\documentclass[10pt,conference]{IEEEtran}

\usepackage{nicefrac}
\usepackage{siunitx}
\usepackage{array,framed}
\usepackage{booktabs}
\usepackage{
  color,
  float,
  epsfig,
  wrapfig,
  graphics,
  graphicx,
  subcaption
}
\usepackage{textcomp,amssymb}
\usepackage{setspace}
\usepackage{latexsym,fancyhdr,url}
\usepackage{enumerate}
\usepackage{algorithm2e}
\usepackage{algpseudocode}
\usepackage{graphics}
\usepackage{xparse} 
\usepackage{xspace}
\usepackage{multirow}
\usepackage{csvsimple}
\usepackage{balance}
\usepackage{amsmath,eqparbox,booktabs}
\usepackage{threeparttable}

\usepackage{
  tikz,
  pgfplots,
  pgfplotstable
}
\usepackage{hyperref}

\usetikzlibrary{
  shapes.geometric,
  arrows,
  external,
  pgfplots.groupplots,
  matrix
}

\usepackage{cite}

\ifCLASSINFOpdf
\else
\fi
\usepackage{amsmath}

\IEEEoverridecommandlockouts
\IEEEpubid{\makebox[\columnwidth]{978-1-7281-5684-2/20/\$31.00~\copyright{}2021 IEEE \hfill} \hspace{\columnsep}\makebox[\columnwidth]{ }}

\begin{document}
%
\title{Detecting Attacks on IoT Devices\\using Featureless 1D-CNN}

\author{\IEEEauthorblockN{Arshiya Khan}
\IEEEauthorblockA{University of Delaware\\
Newark, Delaware, USA\\
Email: arshiyak@udel.edu\\
}
\and
\IEEEauthorblockN{Chase Cotton}
\IEEEauthorblockA{University of Delaware\\
Newark, Delaware, USA\\
Email: ccotton@udel.edu\\
}}


\maketitle

\begin{abstract}
The generalization of deep learning has helped us, in the past, address challenges such as malware identification and anomaly detection in the network security domain. However, as effective as it is, scarcity of memory and processing power makes it difficult to perform these tasks in Internet of Things (IoT) devices. This research finds an easy way out of this bottleneck by depreciating the need for feature engineering and subsequent processing in machine learning techniques. In this study, we introduce a Featureless machine learning process to perform anomaly detection. It uses unprocessed byte streams of packets as training data. Featureless machine learning enables a low cost and low memory time-series analysis of network traffic. It benefits from eliminating the significant investment in subject matter experts and the time required for feature engineering.

\end{abstract}


%
\IEEEpeerreviewmaketitle

\section{Introduction}
\label{sec:intro}

Cybersecurity experts have found insightful features in packet captures, and data scientists have fashioned ideal models out of them that can differentiate malicious traffic from benign traffic \cite{delucia}. Increasingly, fractions of network traffic packets are flowing over secure channels, protected by encryption. Due to this, creating meaningful features for machine learning (ML) has become increasingly complex and will eventually become obsolete. Scientists around the world are now using raw packet data to analyze the traffic \cite{Wang:2017ieee}. The possibility of raw packet analysis has enabled other fields to open their doors for ML technologies. Internet of Things (IoT) is one such field. IoT devices are sensors exposed to the environment, exchanging information over the internet and conditionally reacting to change. Appliances like smart bulbs, voice assistants like Amazon Echo, and any other device that can be controlled remotely are part of IoT. Information getting into and out of these devices passes through several other network devices in the form of packets.\par
Most of these network devices have Intrusion Prevention Systems (IPS), or Intrusion Detection Systems (IDS) installed, which inspect the traffic and mark them as secure or insecure according to their rule-books and detection algorithms. These systems need additional processing power and memory to record observed traffic and trace anomalies at higher speeds. Several studies \cite{nbaIoT,gilberto} have used ML to detect traffic anomalies. Traditionally they use feature-based models. As mentioned earlier, this approach is losing its usability over encrypted channels. It also makes the models less efficient and costly to implement, especially in small, low-powered IoT devices.

\section{Background and Related Work}
\label{sec:relwork}
A variety of exciting research has been done in recent years on network traffic classification, some of which have been summarized in Table~\ref{tab:literature}. Feature-based traffic classification models have used these approaches: 1) port-based, 2) deep packets inspection (DPI)-based, and 3) behavior-based\cite{Velan:2015}.\par
In a \textbf{port-based} approach, reliable ports are classified as benign while the rest, malicious. However, this approach can easily be defeated by port hiding techniques like port camouflaging and randomization.\par
In a \textbf{DPI-based} approach, both the headers and the payload are examined to perform classification. The header provides information about the application which sent the packet, and payload signatures are cross-referenced against known bad signatures for malware identification. This method is inefficient due to its high computational power and time.\par
Lastly, in a \textbf{behavior-based} approach, trends of a flow or session are observed. Statistics of these trends are curated into features.\par
\begin{table*}
    \caption{Few Related Work and Comparison}
    \label{tab:literature}
\begin{center}
\begin{tabular}{ccccccc}
  \toprule
  \textbf{Work} & \textbf{DL Technique} & \textbf{Model Input}  & \textbf{Dataset} & \textbf{Classification Task} &
  \textbf{Accuracy} &
  \textbf{Year}\\
  \midrule
    Wei Wang\cite{Wang:2017} &
    2D-CNN &
    images &
    USTC‐TFC2016\cite{Wang:2017} &
    malicious traffic &
    99.23\% &
    2017 \\

    Wang\cite{Wang:2017ieee} &
    1D-CNN &
    bytes &
    ISCX\cite{icsxvpn2016} &
    traffic characterization &
    100\% &
    2017\\
    
    Lotfollahi\cite{Lotfollahi:2017} &
    1D-CNN+SAE &
    bytes &
    ISCX\cite{icsxvpn2016} &
    traffic characterization &
    98\% &
    2017\\

    Rezaei \& Liu\cite{Rezaei:2020} &
    1D-CNN &
    time series &
    ISCX\cite{icsxvpn2016} \& QUIC\cite{quic} &
    traffic characterization &
    94.67\% &
    2020\\
    
    Marin\cite{Marin:2020} &
    1D-CNN+LSTM &
    raw bytes &
    USTC-TFC2016\cite{Wang:2017} &
    malicious traffic &
    98.6\% &
    2020\\

    \bottomrule
    \end{tabular}

\end{center}
\end{table*}
Velan et al. \cite{Velan:2015} in 2015 published an extensive survey of research studies on encrypted traffic classification between 2005 and 2014. Out of the 26 papers reviewed, 12 used flow features, 5 used packet features, 7 used a combination of flow and packet features, while 2 used other features.\par
An insightful survey was done by Wang et al. \cite{Wang:2019} in 2019 showed a rise in the use of raw packet content in deep learning models. Between 2015 and 2018, 8 out of 12 researchers reported on encrypted traffic classification using raw data from the packet; 2 used both packet features and raw data, while the remaining 2 used only packet-based features and only flow-based features each.\par
Wei Wang et al. \cite{Wang:2017} in 2017 is one of the papers which used flows and sessions collected in the form of raw bytes. This study used representation learning to perform malware traffic classification tasks by converting raw bytes into images. USTC-TFC2016 \cite{Wang:2017}, a private dataset extracted from The Stratosphere IPS Project Malware Dataset \cite{stratodatasets}, was used to perform multi-class classification. The raw bytes collected from packets were converted into images and used as a training dataset for a two-dimensional convolutional neural network (2D-CNN) image classifier, a common machine learning preprocessing step. No feature extraction was required in this study. Experimental results for binary classification (malware or benign) show 100\% accuracy, and for the 10-class and 20-class classification, the model got 99.23\% and 99.17\% accuracy, respectively. The average accuracy of classifiers was 99.41\%. The highest precision for the 10-class experiment was 100\%, and the lowest was 90.7\%, while the recall percentage ranged from 91.1\% to 100\%.\par
A similar study was conducted by Wang et al. \cite{Wang:2017ieee} 2017, where the ISCX VPN-nonVPN \cite{icsxvpn2016} dataset was used to perform encrypted traffic classification. Raw bytes represented in flows and sessions were used to perform multi-class classification using a one-dimensional convolutional neural network (1D-CNN) and 2D-CNN. 1D-CNN performed better or equal in all cases. With binary classification, they achieved 100\% precision; however, for 6-class and 12-class classifications, it was 85.5\% and 85.8\%, respectively.\par
In 2019, Lotfollahi et al. \cite{Lotfollahi:2017} introduced DeepPacket, a deep learning framework to perform traffic characterization and application identification. It comprised two deep learning methods, stacked autoencoder (SAE) neural network and 1D-CNN. ISCX VPN-nonVPN \cite{icsxvpn2016} was used here too. The traffic characterization task gave an average precision of 94\% and an average recall of 93\% with 1D-CNN, and an average precision of 92\%, and an average recall of 92\% with SAE. On the other hand, the application identification task gave an average precision of 98\% and an average recall of 98\% with 1D-CNN, and an average precision of 96\%, and an average recall of 95\% with SAE.\par
Huang et al. \cite{Huang:2019} presented a consolidated study on deep learning for use-cases in the time series domain. In addition to health, finance, transportation, cybersecurity was one of the emerging real-world applications in the time series domain. Common topics include malware identification, traffic classification, and anomaly detection.\par
Rezaei and Liu \cite{Rezaei:2020}, in 2020, reformulated the problem of network traffic classification into two tasks: bandwidth requirement and duration of a flow prediction using multi-task learning. Two tasks were solved for the cost of one. The ISCX VPN-nonVPN \cite{icsxvpn2016} and the QUIC datasets \cite{quic} were used in this study. A 1D-CNN was trained on three time series features, packet length, inter-arrival time, and direction. On the QUIC dataset \cite{quic}, the traffic classification task achieved the highest accuracy of 94.67\%, for which bandwidth requirement and duration of flow achieved 90.67\% and 91.33\%, respectively. On the ISCX dataset \cite{icsxvpn2016}, the traffic classification task achieved the highest accuracy of 80.67\%, for which bandwidth requirement and duration of flow achieved 88.67\% and 90.00\%, respectively.\par
Marin et al. \cite{Marin:2020}, in 2020, introduced two types of malware traffic detection methods, raw packets, and raw flows, to eliminate the need for expert information to engineer features. They used raw byte-stream from pcap (packet capture) files. Each packet is a training instance in a deep learning model based on a combination of a 1D-CNN and a Long short-term memory (LSTM) network. With 98.6\% accuracy, a flow experiment performed better than a packet experiment which only achieved 77.6\% accuracy. A random forest (RF) trained on 200 in-flow features was tested against both methods to test the performance of a feature-less model against traditional models with features designed by domain experts. Raw packets and flows both performed better than RF both in terms of accuracy and false-positive rates.\par
As we can see, gradually, the task of malware detection has been moving from featured ML to direct ML. However, the studies above have not considered the applicability of their models to all environments. LSTMs and 2D-CNNs are practically impossible to use on small IoT devices. They need numerous preprocessing steps to implement features and train a model on them.\par

\section{Methodology}
\label{sec:methodology}
\subsection{Our approach}
\label{sec:our approach}

ML-based anomaly detection techniques have shown promising results in the past \cite{limthong,zenati}. Malware detection in network traffic is necessary nonetheless an overhead in customer-facing services like IoT devices. IoT devices have a small window to validate incoming data and have a limited battery life to sustain prolonged operations. Therefore, implementing ML-based malware detection models is not realistic, as they involve heavy feature engineering and need deeper, complex models for training. Feature engineering comprises defining features, extracting them from raw data, cleaning, scaling, and consolidating. It is an intricate and time-consuming instantaneous process. In this study, we have skipped this cumbersome process and introduced an IoT-friendly methodology. We call this approach featureless modeling.\par
\textbf{Featureless modeling} is becoming a necessity of a secure IoT ecosystem. In featureless modeling, we do not manipulate the stream of raw packets arriving at a device. Instead, we convert it into a byte stream which constitutes the training set of our ML model. Consequently, the feature extraction process is removed from the ML pipeline.\par

Two major overheads are eliminated using featureless modeling technique: \textit{computation cost} and \textit{human cost}. Computation cost covers the computational overhead that a model suffers from completing feature engineering. It can involve data collection, cropping, and cleaning. Human cost involves the domain knowledge needed to extract features and technical expertise to manipulate them into meaningful predictors.\par
We have divided our traffic classification study into three representations: 1) Raw session-based, 2) Raw flow-based, and 3) Raw packet-based. We have used all three different representations of a dataset for three unique ML models. Finally, we have compared model performance to determine the best one.\par
\textbf{Raw session-based:} A session is a bi-directional dialogue between two IoT devices. All packets in a session use the same 5-tuple information: source IP (Internet Protocol) address, source port, destination IP address, destination port, and transport protocol, present in each packet header. In this representation of the experiment, we split a pcap (packet capture) file into individual sessions using \textit{pcapsplitter}\cite{pcapsplitter} tool created by MIT. One session represents a sample for the ground truth.\par
\textbf{Raw flow-based:} A flow is a uni-directional instance of communication from one device to another. A flow may contain more than one packet. In this representation, we prepared our dataset by splitting the same pcap file into flows using a tool called \textit{SplitCap}\cite{splitcap}. A byte stream of flow represents a unique sample in the ground truth.\par
\textbf{Raw packet-based:} A packet is an individual unit derived from a live traffic stream recorded in a pcap format file. In this representation of the experiment, we recorded the byte stream of each packet and collected it into a row of byte values. A packet is a sample in the dataset.\par
A study published in 2020 has also implemented a similar technique where they used raw bytes \cite{Marin:2020}. However, they have not demonstrated that their technique works faster due to the elimination of features. In addition, their model contains both 1D-CNN and LSTM layers which can increase the complexity of  a model and result in over-fitting.\par
Our contribution is to introduce a network traffic classification system, which is much faster and consumes less memory. We will achieve this by implementing featureless modeling and using smaller models.

\subsection{Experimental design}
\label{sec:experimentaldesign}

We have performed experiments for this study on the Aposemat IoT-23 dataset \cite{iot23}. It is a labeled network traffic dataset captured between 2018-2019 from three IoT devices. It was published in 2020 by Stratosphere Research Lab. It contains both original pcap files and Zeek files of the captured traffic. Only a few studies have mentioned it in their work. Bobravnikova et al. \cite{bobravnikova} performed botnet detection using this dataset. However, they extracted numerical feature vectors for their classification task and achieved 98\% accuracy on support vector machines (SVM) \cite{Cortes1995SupportvectorN}. Stoian \cite{stoian} used features from conn.log.labeled file generated by Zeek on this dataset. They also performed multi-label classification and achieved 100\% accuracy with a Random Forest classifier \cite{10.5555/844379.844681}.\par
IoT-23 has three benign and 20 malicious scenarios. We collected benign traffic from three IoT devices: Amazon echo, Somfy smart door lock, and Philips hue LED lamp. The malicious data was collected by infecting these IoT devices with a variety of botnet attacks in a simulated environment.\par
\begin{figure}[htbp!]
  \centering
  \includegraphics[width=\linewidth]{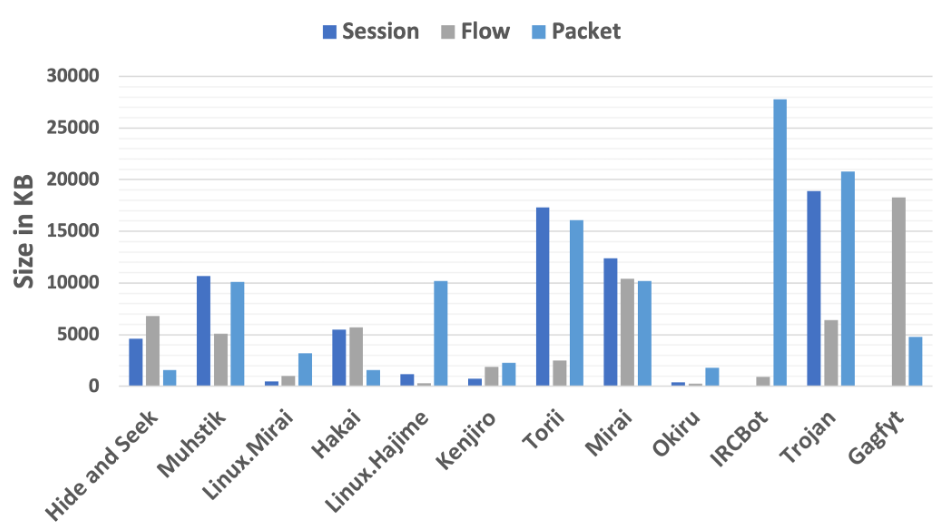}
  \caption{Byte Distribution}
  \label{fig:bytedistribution}
\end{figure}
Since our study focuses on a featureless model, we have selected raw pcap files of the dataset. We have taken all three benign scenarios and integrated them into one to increase the size of the training set. The malicious scenarios are pcap files recorded at devices infected by 20 botnet attacks. Out of the 20 malicious scenarios, some were repetitive. We have eliminated the redundancy and used 12 unique scenarios. They are: 1) Hide and Seek, 2) Muhstik, 3) Linux.Mirai, 4) Hakai, 5) Linux.Hajime, 6) Kenjiro, 7) Torii, 8) Mirai, 9) Okiru, 10) IRCBot, 11) Trojan, and 12) Gagfyt. Figure~\ref{fig:bytedistribution} displays the byte distribution of infected traffic across the 12 botnet types in our training set. We have performed two sets of experiments: 1) Binary classification between benign and malicious traffic and 2) multi-label classification between the 12 botnet traffic.\par
As mentioned in section \ref{sec:our approach}, we have represented this dataset into three different views: session, flow, and packet.\par
The \textit{session view}  is composed of unique sessions. We have used tshark \cite{tshark} to convert the byte stream of a session to the hexadecimal format. Each byte stream forms a row in the training data.\par
The \textit{flow view} uses the same technique as above to extract a hexadecimal stream of each flow.\par
The \textit{packet view} uses hexadecimal representations of packets as a sample in the labeled training set.\par
Based on these views, we created three different training sets and trained them separately. Analysis of experimental results from all three views will help us use this technique in IoT devices and provide maximum security from botnet attacks.\par
\begin{figure}[htbp!]
\centering
  \includegraphics[width=\columnwidth]{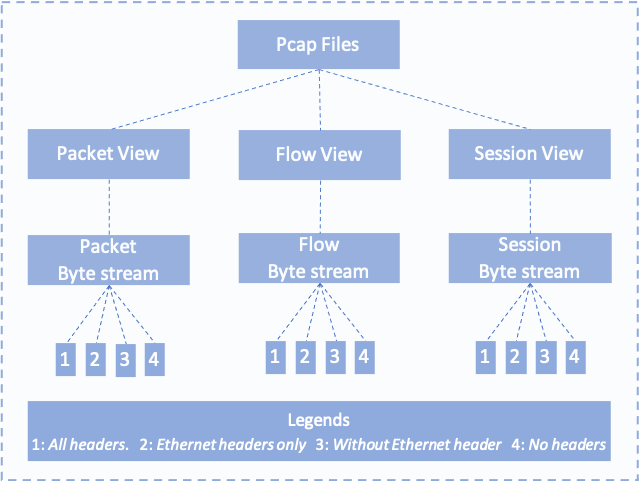}
  \caption{Training data generation}
  \label{fig:trainingsetgeneration}
\end{figure}
To further achieve our goal of reducing computational overhead, we have experimented with Ethernet and IP headers in the packet. We divide the above views into four unique categories. In the first category, the whole packet is included, which means the training data contains both Ethernet and IP headers, and we call this category \textit{All headers}. We call the second category: \textit{Ethernet headers only} as we keep ethernet headers and remove IP headers from the training dataset. Intuitively, in the third category, ethernet headers are removed from the packet, so we call this category \textit{Without ethernet}. In the last category, we removed both ethernet and IP headers, so it is called the \textit{No headers} category. All the three views (session, flow, and packet) are further branched into these four categories and then trained separately using featureless modeling. As shown in Figure~\ref{fig:trainingsetgeneration}, there are a total of 12 different experiments in our study.
\begin{table}[htbp]
\caption{Model}
\label{tab:model}
\begin{center}
\begin{tabular}{|c|c|c|c|} 
 \hline
 \textbf{Layer} & \textbf{Operation}  & \textbf{Filter}  & \textbf{Output}\\
 \hline
0 & conv1d &  64*3 & 18*64\\ 

 \hline
1 & max\_pool1d & 5*1 & 3*64\\ 

 \hline
2 & conv1d & 64*3 & 1*64 \\ 

 \hline
3 & global\_avg\_pool1d & - & 64 \\ 

 \hline
4 & dense & - & 2/12 \\

 \hline
 \end{tabular}
\end{center}

\end{table}

\subsection{DL architecture}
\label{sec:dl architecture}

Packets incoming on a device are instances of data distributed in time, making this a time series problem. In the past, 1D-CNNs have been used to classify time series data \cite{10.1007/978-3-642-35395-6_30}. They have also been used in Natural Language Processing (NLP) models \cite{kim-2014-convolutional}. In this study, we have developed a small 1D-CNN that can be implemented in smaller devices and detect malicious streams of packets.\par
Our neural network, as shown in Table \ref{tab:model}, consists of two 1D-CNN layers, a maxpooling layer, a global average pooling layer, and a fully connected Dense layer in the end. The first convolution layer performs convolution with a kernel size of 64 and a stride of 3 on the input vectorized byte stream. The second convolution layer also has the same parameters. The maxpooling layer lies between the two convolution layers and performs a 5*1 pooling operation to reduce the output size from the first layer. We have used a global average pooling layer to reduce model over-fitting. In the end, we used a Dense layer which is used to learn non-linear relationship between model weights. We have used binary cross-entropy as a loss function for binary classification and softmax function for activation. For categorical classification, we have used categorical cross-entropy as a loss function and sigmoid function as activation. While the binary classifier classifies malicious and benign traffic, the multi-label classifier performs classification between 12 botnet classes. The model used a batch size of 20 and trained for 50 epochs using Keras \cite{chollet2015keras}.It also used a checkpoint function to find the best model among all epochs. Using checkpoint enables TensorFlow \cite{tensorflow2015-whitepaper} to stop training when it achieves the maximum value of the evaluation metric, which is accuracy in our case. Then, we saved the hyperparameter weights of that epoch to a file that we can keep on IoT devices to perform classification on demand. Hyperparameters that are not optimal result in a complex model, which leads to over-fitting and high validation loss.\par 
We used an Nvidia GeForce GTX 1060 GPU for training on a 12GB Ubuntu 16.04 server on an x86 architecture.

\section{Evaluation}
\label{sec:eval}
\subsection{Evaluation Metrics}
\label{sec:evalMetrics}
\begin{figure}[htbp!]
  \centering
  \includegraphics[width=\linewidth]{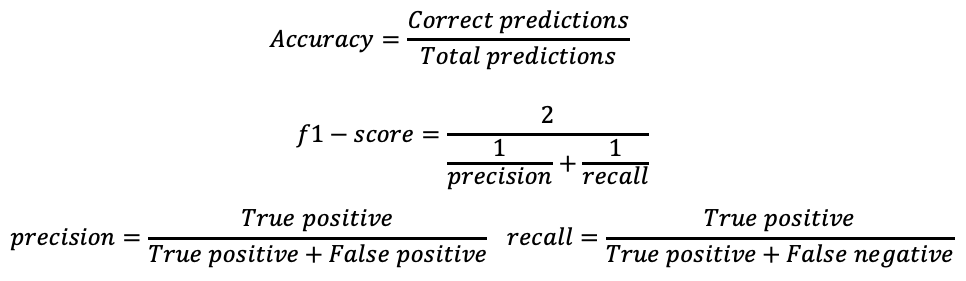}
  \caption{Evaluation Metrics}
  \label{fig:evaluation}
\end{figure}

We evaluated the experiments on two metrics: accuracy and f-1 score, as shown in Figure~\ref{fig:evaluation}. Accuracy is calculated on both training data and validation data. It is defined as the percentage of correct classification out of total classifications performed by the model. The f-1 score is also a metric of the model's performance. It is the weighted average of precision and recall values on the validation data.\par

A series of detailed experiments indicate that binary classification between malicious and benign traffic gave the best results on this dataset. As displayed in Table~\ref{tab:binaryresults}, binary classification achieved 100\% accuracy in each header category of the session view. It also achieved 100\% accuracy in flow view when all headers were included. Packet view achieved maximum accuracy of 100\% when all headers were included. Flow view achieved the highest f-1 scores in the binary classification. It recorded a 100\% f-1 score when no headers were included. Session view achieved 98\% f-1 score when only IP header was included. Packet view achieved 97\% with the same header setting.\par
\begin{table}[h!]
\caption{Binary Classification Performance}
\label{tab:binaryresults}
\begin{center}
\begin{tabular}{|c|c|c|c|} 
\hline
\textbf{View} &\textbf{Header}& \textbf{Accuracy}& \textbf{f-1 score} \\
  \hline
  \multirow{4}{3em}{Session} & All headers & 1.00 & 0.97\\ 
  & Only ethernet & 1.00 & 0.96\\ 
  & Without ethernet & 1.00 & 0.98\\
  & No headers & 1.00 & 0.94\\
  \hline
  \multirow{4}{3em}{Flow} & All headers & 1.00 & 0.97\\ 
  & Only ethernet & 1.00 & 0.93\\ 
  & Without ethernet & 0.97 & 0.96\\
  & No headers & 0.99 & 1.00\\ 
  \hline
  \multirow{4}{3em}{Packet} & All headers & 1.00 & 0.97\\ 
  & Only ethernet & 0.98 & 0.96\\ 
  & Without ethernet & 0.98 & 0.97\\
  & No headers & 0.99 & 0.95\\ 
  \hline
\end{tabular}
\end{center}
\end{table}

Results of the multi-label classification between 12 botnet classes are detailed in Table~\ref{tab:multilabelresults}. It achieved a maximum accuracy of 99\% in all views. In session experiments, f-1 scores were highest 86\% when all headers were included and lowest 79\% when all headers were removed. However, in flow experiments, the f-1 score was the highest, 88\% in all headers category. However, it was the lowest when either of the headers was removed, 85\%. f-1 score recovered when both headers were removed to 87\%, which is only 0.1 \% lower than all headers category. Packet view saw the same trend as flow. f-1 score dipped by when either header was removed from the dataset.\par
\begin{table}[h!]
\caption{Multi-Label Classification Performance}
\label{tab:multilabelresults}
\begin{center}
\begin{tabular}{|c|c|c|c|} 
\hline
\textbf{View} &\textbf{Header}& \textbf{Accuracy}& \textbf{f-1 score} \\
  \hline
  \multirow{4}{3em}{Session} & All headers & 0.99 & 0.86\\ 
  & Only ethernet & 0.99 & 0.84\\ 
  & Without ethernet & 0.99 & 0.85\\
  & No headers & 0.94 & 0.79\\
  \hline
  \multirow{4}{3em}{Flow} & All headers & 0.99 & 0.88\\ 
  & Only ethernet & 0.99 & 0.85\\ 
  & Without ethernet & 0.99 & 0.85\\
  & No headers & 0.99 & 0.87\\ 
  \hline
  \multirow{4}{3em}{Packet} & All headers & 0.99 & 0.84\\ 
  & Only ethernet & 0.96 & 0.82\\ 
  & Without ethernet & 0.93 & 0.82\\
  & No headers & 0.99 & 0.83\\ 
  \hline
\end{tabular}
\end{center}

\end{table}
\begin{figure}[htbp!]
  \centering
  \includegraphics[width=\linewidth]{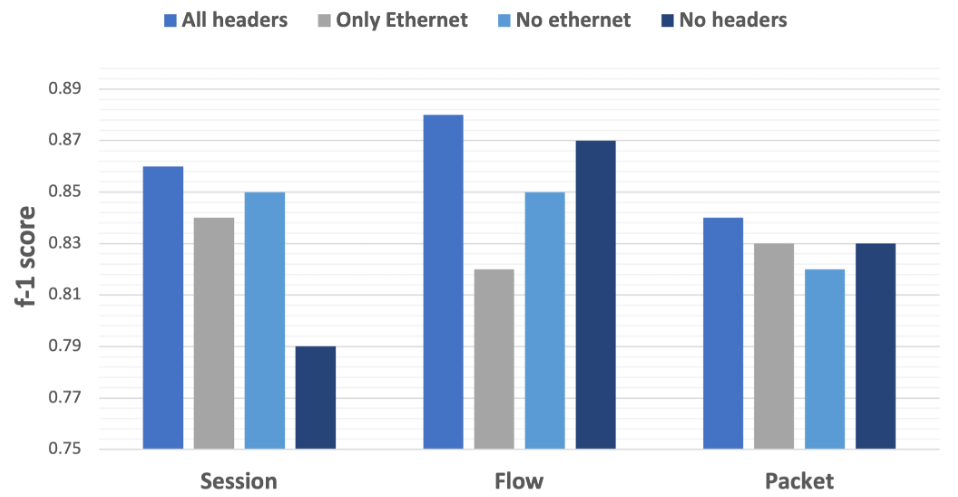}
  \caption{f-1 score trends in all 3 views}
  \label{fig:f1score}
\end{figure}
Accuracy was highest when all headers were included, reduced when either of the headers was removed but recovered when there were no headers. f-1 score was highest when all headers were included, decreased when either header was removed from the dataset. The f-1 score decreased further when all headers were removed from the training set with a few exceptions.\par
From this study, we can safely deduce that headers impact heavily on the decision-making capability of anomaly detection algorithms. Contradictory to the norm of cropping them out from the training set \cite{Marin:2020}, better results are achieved when we include them as indicated in Figure \ref{fig:f1score}. This architecture of a featureless model is more synchronous with the theme of machine learning being able to make decisions of working in an unknown environment and perform on data as they come.

\subsection{Performance on IoT}
\label{sec:perf_iot}

To be deployed on IoT devices, an ML model should be fast. In this section, we have compared the performance of our featureless model with another feature-based model. For our featureless model, we used \textit{all headers} category. For the feature-based model, we used a malware detection model called N-BaIoT published in 2018\cite{nbaIoT}. This model had 115 numerical features, and it was trained on autoencoder models for detecting malware on IoT devices. We performed binary classification between malicious and benign traffic on both models and trained and tested them on the same underlying architecture.\par
\begin{table*}
    \caption{Performance for IoT devices}
    \label{tab:iot_performance}
\begin{center}
\begin{tabular}{cccc}
  \toprule
\textbf{Binary classification}& \textbf{Accuracy} &\textbf{Training time (sec)}& \textbf{Testing time (sec)} \\
\midrule
    Session & 100\% & 2.813 & 1.171\\ 
    Flow & 100\% & 2.161 & 0.513\\ 
    Packet & 100\% & 15.502 & 1.350 \\ 
    N-BaIoT & 99.95\% & 20.763 & 1.238\\ 
    \bottomrule
    \end{tabular}

\end{center}
\end{table*}
We used the training and testing time of the model as performance metrics. To measure these parameters, we used the \textit{time} function in Python. Table \ref{tab:iot_performance} shows that the training time in all three representations of the featureless model is significantly lower than the feature-based model. The testing time of packet view is higher than N-BaIoT by 0.112 seconds. This comparison indicates that featureless models are much more efficient than feature-based models, and they can save a significant amount of time and power.\par

\section{Conclusion}

Previous works in the area of this study have discarded headers from their training set. However, in our experiment, we have compared the performance of the same dataset with and without headers. Results from this extensive experimentation solidify our claim that preprocessing and removal of headers from the packet is an unnecessary step in the traffic classification task. Handicapped by running on limited battery power, IoT devices need to find new technologies with less computational overhead and more accurate results. It is possible by using featureless modeling with 1D-CNNs. Not only can this methodology identify anomalies in traffic with almost 100\% accuracy, but it also takes lesser time to train and test the models. In the future, this research will attempt to find smaller and more effective IoT-friendly neural networks that can be deployed to live IoT devices.





\bibliographystyle{IEEEtran}
\bibliography{bib}

\begin{thebibliography}{10}
\providecommand{\url}[1]{#1}
\csname url@samestyle\endcsname
\providecommand{\newblock}{\relax}
\providecommand{\bibinfo}[2]{#2}
\providecommand{\BIBentrySTDinterwordspacing}{\spaceskip=0pt\relax}
\providecommand{\BIBentryALTinterwordstretchfactor}{4}
\providecommand{\BIBentryALTinterwordspacing}{\spaceskip=\fontdimen2\font plus
\BIBentryALTinterwordstretchfactor\fontdimen3\font minus
  \fontdimen4\font\relax}
\providecommand{\BIBforeignlanguage}[2]{{%
\expandafter\ifx\csname l@#1\endcsname\relax
\typeout{** WARNING: IEEEtran.bst: No hyphenation pattern has been}%
\typeout{** loaded for the language `#1'. Using the pattern for}%
\typeout{** the default language instead.}%
\else
\language=\csname l@#1\endcsname
\fi
#2}}
\providecommand{\BIBdecl}{\relax}
\BIBdecl

\bibitem{delucia}
M.~De~Lucia and C.~Cotton, ``Importance of features in adversarial machine
  learning for cyber security,'' 11 2018.

\bibitem{Wang:2017ieee}
W.~{Wang}, M.~{Zhu}, J.~{Wang}, X.~{Zeng}, and Z.~{Yang}, ``End-to-end
  encrypted traffic classification with one-dimensional convolution neural
  networks,'' in \emph{2017 IEEE International Conference on Intelligence and
  Security Informatics (ISI)}, 2017, pp. 43--48.

\bibitem{nbaIoT}
Y.~{Meidan}, M.~{Bohadana}, Y.~{Mathov}, Y.~{Mirsky}, A.~{Shabtai},
  D.~{Breitenbacher}, and Y.~{Elovici}, ``N-baiot—network-based detection of
  iot botnet attacks using deep autoencoders,'' \emph{IEEE Pervasive
  Computing}, vol.~17, no.~3, pp. 12--22, 2018.

\bibitem{gilberto}
G.~Junior, J.~Rodrigues, L.~Carvalho, J.~Al-Muhtadi, and M.~Proença, ``A
  comprehensive survey on network anomaly detection,'' \emph{Telecommunication
  Systems}, vol.~70, 07 2018.

\bibitem{Velan:2015}
M.~Velan, Petr an d~\v{C}erm\'{a}k, P.~\v{C}eleda, and M.~Dra\v{s}ar, ``A
  survey of methods for encrypted traffic classification and analysis,''
  \emph{Netw.}, vol.~25, no.~5, p. 355–374, Sep. 2015.

\bibitem{Wang:2017}
{Wei Wang}, {Ming Zhu}, {Xuewen Zeng}, {Xiaozhou Ye}, and {Yiqiang Sheng},
  ``Malware traffic classification using convolutional neural network for
  representation learning,'' in \emph{2017 International Conference on
  Information Networking (ICOIN)}, 2017, pp. 712--717.

\bibitem{icsxvpn2016}
C.~I. f.~C. UNB, ``Icsx vpn-nonvpn dataset,'' 2016,
  \url{https://www.unb.ca/cic/datasets/vpn.html}.

\bibitem{Lotfollahi:2017}
M.~Lotfollahi, R.~Shirali~hossein zade, M.~Jafari~Siavoshani, and M.~Saberian,
  ``Deep packet: A novel approach for encrypted traffic classification using
  deep learning,'' \emph{Soft Computing}, vol.~24, 09 2017.

\bibitem{Rezaei:2020}
S.~{Rezaei} and X.~{Liu}, ``Multitask learning for network traffic
  classification,'' in \emph{2020 29th International Conference on Computer
  Communications and Networks (ICCCN)}, 2020, pp. 1--9.

\bibitem{quic}
V.~{Tong}, H.~A. {Tran}, S.~{Souihi}, and A.~{Mellouk}, ``A novel quic traffic
  classifier based on convolutional neural networks,'' in \emph{2018 IEEE
  Global Communications Conference (GLOBECOM)}, 2018, pp. 1--6.

\bibitem{Marin:2020}
G.~Marín, P.~Casas, and G.~Capdehourat, ``Deepmal -- deep learning models for
  malware traffic detection and classification,'' 2020.

\bibitem{Wang:2019}
P.~Wang, X.~Chen, F.~Ye, and S.~Zhixin, ``A survey of techniques for mobile
  service encrypted traffic classification using deep learning,'' \emph{IEEE
  Access}, vol.~PP, pp. 1--1, 04 2019.

\bibitem{stratodatasets}
Stratosphere, ``Stratosphere laboratory datasets,'' 2015, retrieved March 13,
  2020, from \url{https://www.stratosphereips.org/datasets-overview}.

\bibitem{Huang:2019}
X.~{Huang}, G.~C. {Fox}, S.~{Serebryakov}, A.~{Mohan}, P.~{Morkisz}, and
  D.~{Dutta}, ``Benchmarking deep learning for time series: Challenges and
  directions,'' in \emph{2019 IEEE International Conference on Big Data (Big
  Data)}, 2019, pp. 5679--5682.

\bibitem{limthong}
K.~{Limthong} and T.~{Tawsook}, ``Network traffic anomaly detection using
  machine learning approaches,'' in \emph{2012 IEEE Network Operations and
  Management Symposium}, 2012, pp. 542--545.

\bibitem{zenati}
H.~{Zenati}, M.~{Romain}, C.~{Foo}, B.~{Lecouat}, and V.~{Chandrasekhar},
  ``Adversarially learned anomaly detection,'' in \emph{2018 IEEE International
  Conference on Data Mining (ICDM)}, 2018, pp. 727--736.

\bibitem{pcapsplitter}
S.~H.~R. PcapPlusPlus, ``Pcapsplitter,'' 2019,
  \url{https://github.com/shramos/pcap-splitter}.

\bibitem{splitcap}
Netresec, ``Splitcap,'' 2010, used in 2020 \url{https://www.netresec.com/}.

\bibitem{iot23}
M.~J.~E. Agustin~Parmisano, Sebastian~Garcia, ``Stratosphere laboratory
  aposemat iot-23,'' 2020,
  \url{https://www.stratosphereips.org/datasets-iot23}.

\bibitem{bobravnikova}
K.~Bobrovnikova, S.~Lysenko, P.~Gaj, V.~Martynyuk, and D.~Denysiuk, ``Technique
  for iot cyberattacks detection based on dns traffic analysis,'' 2020,
  \url{http://ceur-ws.org/Vol-2623/paper19.pdf}.

\bibitem{Cortes1995SupportvectorN}
C.~Cortes and V.~Vapnik, ``Support-vector networks,'' \emph{Machine Learning},
  vol.~20, pp. 273--297, 1995.

\bibitem{stoian}
\BIBentryALTinterwordspacing
N.~{Stoian}, ``Machine learning for anomaly detection in iot networks : Malware
  analysis on the iot-23 data set,'' July 2020. [Online]. Available:
  \url{http://essay.utwente.nl/81979/}
\BIBentrySTDinterwordspacing

\bibitem{10.5555/844379.844681}
T.~K. Ho, ``Random decision forests,'' in \emph{Proceedings of the Third
  International Conference on Document Analysis and Recognition (Volume 1) -
  Volume 1}.\hskip 1em plus 0.5em minus 0.4em\relax USA: IEEE Computer Society,
  1995, p. 278.

\bibitem{tshark}
Wireshark, ``tshark,'' 2006, used in 2020
  \url{https://www.wireshark.org/docs/man-pages/tshark.html}.

\bibitem{10.1007/978-3-642-35395-6_30}
D.~Anguita, A.~Ghio, L.~Oneto, X.~Parra, and J.~L. Reyes-Ortiz, ``Human
  activity recognition on smartphones using a multiclass hardware-friendly
  support vector machine,'' in \emph{Ambient Assisted Living and Home Care},
  J.~Bravo, R.~Herv{\'a}s, and M.~Rodr{\'i}guez, Eds.\hskip 1em plus 0.5em
  minus 0.4em\relax Berlin, Heidelberg: Springer Berlin Heidelberg, 2012, pp.
  216--223.

\bibitem{kim-2014-convolutional}
\BIBentryALTinterwordspacing
Y.~Kim, ``Convolutional neural networks for sentence classification,'' in
  \emph{Proceedings of the 2014 Conference on Empirical Methods in Natural
  Language Processing ({EMNLP})}.\hskip 1em plus 0.5em minus 0.4em\relax Doha,
  Qatar: Association for Computational Linguistics, Oct. 2014, pp. 1746--1751.
  [Online]. Available: \url{https://www.aclweb.org/anthology/D14-1181}
\BIBentrySTDinterwordspacing

\bibitem{chollet2015keras}
F.~Chollet \emph{et~al.}, ``Keras,'' \url{https://keras.io}, 2015.

\bibitem{tensorflow2015-whitepaper}
\BIBentryALTinterwordspacing
M.~Abadi, A.~Agarwal, P.~Barham, E.~Brevdo, Z.~Chen, C.~Citro, G.~S. Corrado,
  A.~Davis, J.~Dean, M.~Devin, S.~Ghemawat, I.~Goodfellow, A.~Harp, G.~Irving,
  M.~Isard, Y.~Jia, R.~Jozefowicz, L.~Kaiser, M.~Kudlur, J.~Levenberg,
  D.~Man\'{e}, R.~Monga, S.~Moore, D.~Murray, C.~Olah, M.~Schuster, J.~Shlens,
  B.~Steiner, I.~Sutskever, K.~Talwar, P.~Tucker, V.~Vanhoucke, V.~Vasudevan,
  F.~Vi\'{e}gas, O.~Vinyals, P.~Warden, M.~Wattenberg, M.~Wicke, Y.~Yu, and
  X.~Zheng, ``{TensorFlow}: Large-scale machine learning on heterogeneous
  systems,'' 2015, software available from tensorflow.org. [Online]. Available:
  \url{https://www.tensorflow.org/}
\BIBentrySTDinterwordspacing

\end{thebibliography}

%
\end{document}